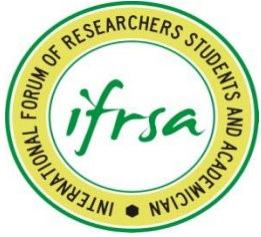

# A Comparative Study of CPU Scheduling Algorithms

Neetu Goel
Research Scholar, TEERTHANKER MAHAVEER UNIVERSITY
Dr. R.B. Garg
Professor
Delhi School of Professional Studies & Research
(Affiliated to GGSIP University, Delhi)

### ABSTRACT

Developing CPU scheduling algorithms and understanding their impact in practice can be difficult and time consuming due to the need to modify and test operating system kernel code and measure the resulting performance on a consistent workload of real applications. As processor is the important resource, CPU scheduling becomes very important in accomplishing the operating system (OS) design goals. The intention should be allowed as many as possible running processes at all time in order to make best use of CPU.

This paper presents a state diagram that depicts the comparative study of various scheduling algorithms for a single CPU and shows which algorithm is best for the particular situation. Using this representation, it becomes much easier to understand what is going on inside the system and why a different set of processes is a candidate for the allocation of the CPU at different time. The objective of the study is to analyze the high efficient CPU scheduler on design of the high quality scheduling algorithms which suits the scheduling goals.

### INTRODUCTION

In a single-processor system, only one process can run at a time; any others must wait until the CPU is free and can be rescheduled. The objective of multiprogramming is to have some process running at all times, to maximize CPU utilization [1]. Scheduling is a fundamental operating-system function. Almost all computer resources are scheduled before use. The CPU is, of course, one of the primary computer resources. Thus, its scheduling is central to operating-system design. CPU scheduling determines which processes run when there are multiple run-able processes. CPU scheduling is important because it can have a big effect on resource utilization and the overall performance of the system [2].

OS may feature up to 3 distinct types of schedulers: a long term scheduler (also known as an admission scheduler or high level scheduler), a mid-term or medium-term scheduler and a short-term scheduler (also known as a dispatcher or CPU scheduler). The rest of the paper is organized as follows: Section II gives review on scheduling algorithms. The proposed algorithm, Results and Discussion have been given in section II

*A. Long-term Scheduler*

The long-term or admission scheduler decides which jobs or processes are to be admitted to the ready queue; that is, when an attempt is made to execute a process its admission to the set of currently executing processes is either authorized or delayed by the long-term scheduler. Thus, this scheduler dictates what processes are to run on a system, and the degree of concurrency to be supported at any one time.

*B. Mid-term Scheduler*

The mid-term scheduler temporarily removes process from main memory and place them on secondary memory (such as a disk drive) or vice versa. This is commonly referred to as "swapping of processes out" or "swapping in" (also incorrectly as "paging out" or "paging in").

*C. Short-term Scheduler*

The short-term scheduler (also known as the CPU scheduler) decides which of processes in the ready queue, in memory are to be executed (allocated a CPU) next following a clock interrupt, an Input-Output (IO) interrupt and an OS call or another form of signal. Thus the short-term scheduler makes scheduling decisions much more frequent than the long-term or mid-term schedulers. This scheduler can be preemptive, implying that it is capable of forcibly removing processes from a CPU when it decides to allocate that CPU to another process, or non pre-emptive (also known as "voluntary" or "co-operative"), in that case the scheduler is unable to force processes off the CPU.

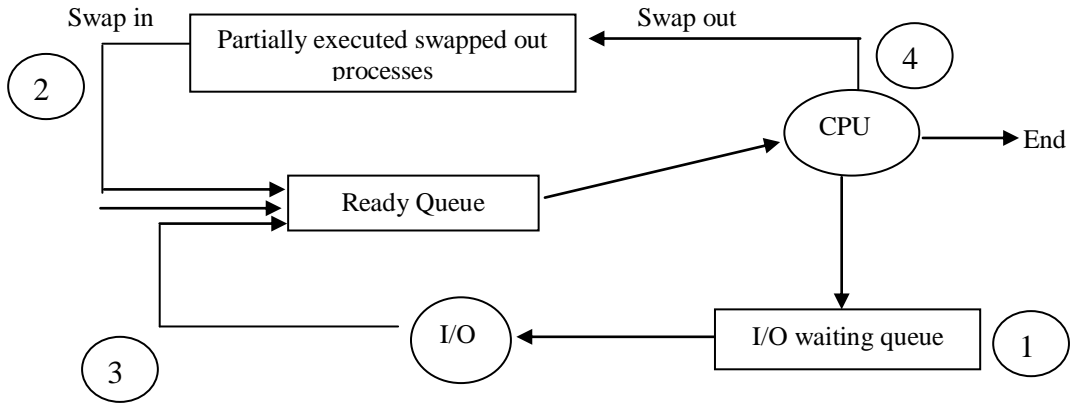

**Figure 1: Shows the following states have been executed in the CPU Scheduler**

1. When a process switches from the running state to the waiting state.
2. When a process switches from the running state to the ready state.
3. When a process switches from the waiting state to the ready state.
4. When a process terminates.

The success of a CPU scheduler depends on the design of high quality scheduling algorithm. High-quality CPU scheduling algorithms rely mainly on criteria such as CPU utilization rate, throughput, turnaround time, waiting time and response time. Thus, the main impetus of this work is to develop a generalized optimum high quality scheduling algorithm suited for all types of job.

## SCHEDULING CRITERIA

Different CPU scheduling algorithms have different properties, and the choice of a particular algorithm may favor one class of processes over another. In choosing which algorithm to use in a particular situation, we must consider the properties of the various algorithms. Many criteria have been suggested for comparing CPU scheduling algorithms. Which characteristics are used for comparison can make a substantial difference in which algorithm is judged to be best. The criteria include the following:

1. **Utilization/Efficiency**: keep the CPU busy 100% of the time with useful work
2. **Throughput:** maximize the number of jobs processed per hour.
3. **Turnaround time**: from the time of submission to the time of completion, minimize the time batch users must wait for output
4. **Waiting time**: Sum of times spent in ready queue - Minimize this
5. **Response Time**: time from submission till the first response is produced, minimize response time for interactive users
6. **Fairness**: make sure each process gets a fair share of the CPU

**CPU Scheduler** whenever the CPU becomes idle; the operating system must select one of the processes in the ready queue to be executed. The selection process is carried out by the short term scheduler (or CPU scheduler). The scheduler selects from among the processes in the memory that are ready to execute and allocates the CPU to one of them Figure 2 shows a schematic of scheduling:

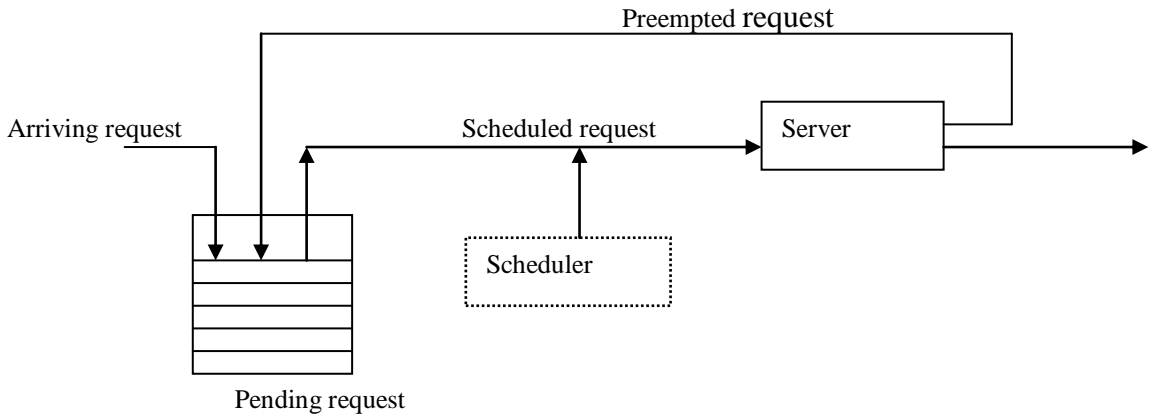

**Figure 2: Schematic of Scheduling**

The ready queue is not necessarily a first-in, first out (FIFO) queue. It may be implemented as a FIFO queue, priority queue, a tree or simply an unordered linked list. Conceptually, however, all the processes in the ready queue are lined up waiting for a chance to run on the CPU. (Stalling William, 2004) An operating system must allocate computer resources among the potentially competing requirements of multiple processes. In the case of the processor, the resource to be allocated is execution time on the processor and the means of allocation is scheduling. This way, the scheduler is the component of the operating system responsible to grant the right to CPU access to a list of several processes ready to execute. This idea is illustrated in the five state diagram of figure 3(Galvin et. al. 2001).

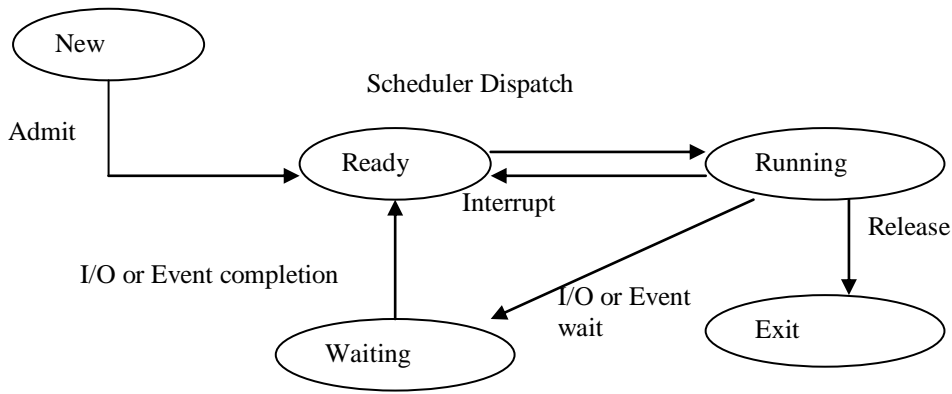

**Figure 3: Life Cycle of Process**

In Circumstances first and fourth, there is no choice in terms of scheduling. A new process (if one exists in the ready queue) must be selected for execution. There is a choice, however, in circumstances second and third. When scheduling takes place under circumstances first and fourth, then the scheduling scheme is non-preemptive; otherwise, the scheduling scheme is preemptive.

Under non-preemptive scheduling, once the CPU has been allocated to a process, the process keeps the CPU until it releases the CPU either by terminating or by switching to the waiting state.

**SCHEDULING ALOGRITHMS**

A. *Algorithm and its characteristics*
   The fundamental scheduling algorithms and its characteristics are described in this section.
a. *First Come First Serve*
   The most intuitive and simplest technique is to allow the first process submitted to run first. This approach is called as first-come, first-served(FCFS) scheduling. In effect, processes are inserted into the tail of a queue when they are submitted[1]. The next process is taken from the head of the queue when each finishes running. This idea is illustrated in the four state diagram of figure 4.

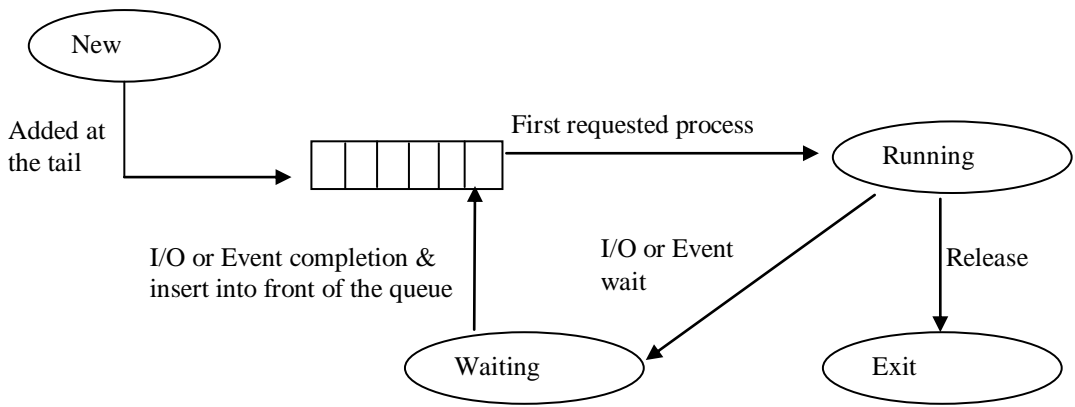

**Figure 4: First Come First Serve Scheduling**

**Characteristics**
- The lack of prioritization does permit every process to eventually complete, hence no starvation.
- Turnaround time, waiting time and response time is high.
- One, Process with longest burst time can monopolize CPU, even if other process burst time is too short. Hence throughput is low [3].

b. *Non preempted Shortest Job First*

The process is allocated to the CPU which has least burst time. A scheduler arranges the processes with the least burst time in head of the queue and longest burst time in tail of the queue. This requires advanced knowledge or estimations about the time required for a process to complete[1]. This algorithm is designed for maximum throughput in most scenarios. This idea is illustrated in the four state diagram of figure 5

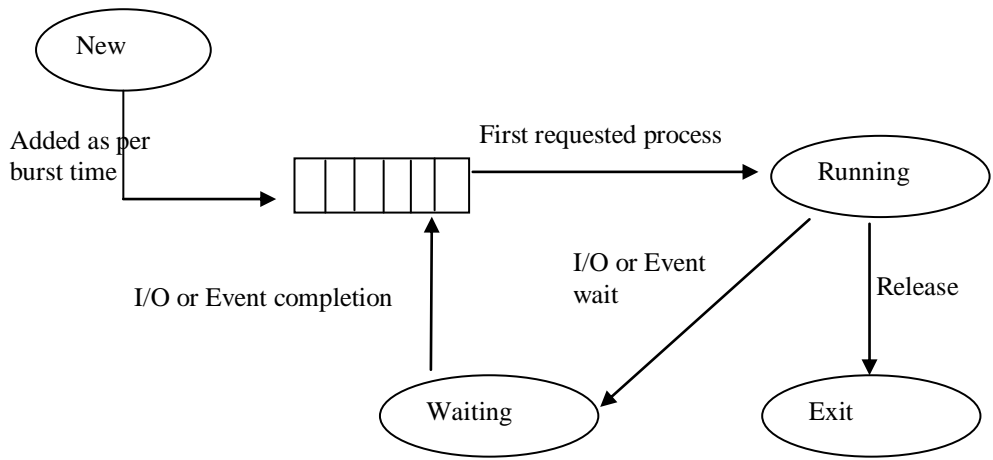

**Figure 5: Shortest Job First Scheduling**

**Characteristics**
- The real difficulty with the SJF algorithm is, to know the length of the next CPU request.
- SJF minimizes the average waiting time[3] because it services small processes before it services large ones. While it minimizes average wait time, it may penalize processes with high service time requests. If the ready list is saturated, then processes with large service times tend to be left in the ready list while small processes receive service. In extreme case, when the system has little idle time, processes with large service time will never be served. This total starvation of large processes is a serious liability of this algorithm.

c. *Round Robin*

The Round Robin (RR) scheduling algorithm assigns a small unit of time, called time slice or quantum. The ready processes are kept in a queue. The scheduler goes around this queue, allocating the CPU to each process for a time interval of assigned quantum. New processes are added to the tail of the queue [4]. This idea is illustrated in the four state diagram of figure 6.

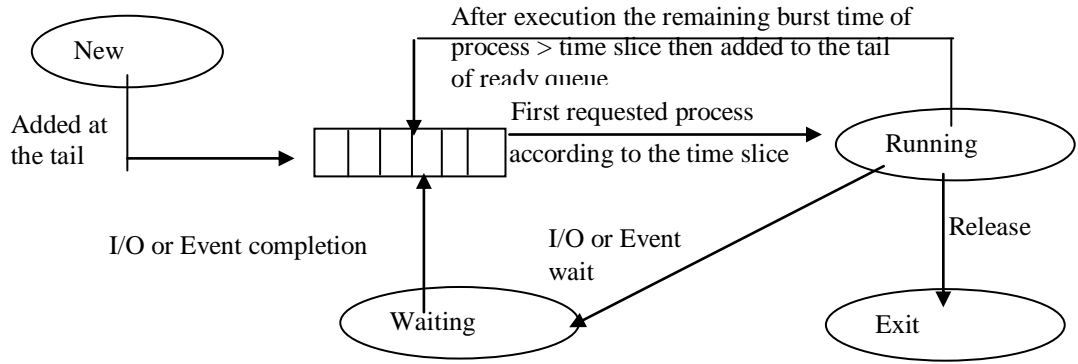

**Figure 6: Round Robin Scheduling**

*Characteristics*

- Setting the quantum too short causes too many context switches and lower the CPU efficiency.
- Setting the quantum too long may cause poor response time and approximates FCFS.
- Because of high waiting times, deadlines are rarely met in a pure RR system.

d. *Priority Scheduling*

The O/S assigns a fixed priority rank to each process. Lower priority processes get interrupted by incoming higher priority processes. This idea is illustrated in the four state diagram of figure 7

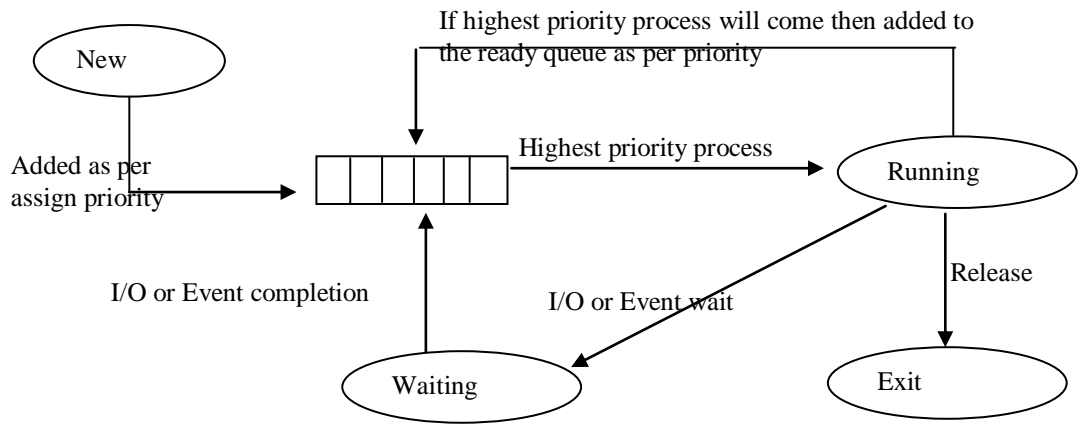

**Figure 7: Priority Scheduling**

*Characteristics*

- Starvation can happen to the low priority process.
- The waiting time gradually increases for the equal priority processes [5].
- Higher priority processes have smaller waiting time and response time.

**B. *Computation of Gantt chart, Waiting Time and Turnaround Time***

Consider the following set of processes, with the length of the CPU-burst time in milliseconds is shown in Table 1:

| Process ID | Burst Time(ms) |
|---|---|
| P0 | 12 |
| P1 | 2 |
| P2 | 3 |
| P3 | 2 |
| P4 | 6 |

**Table 1: Processes with Its Id and Burst Time**

*a. First Come First Serve*

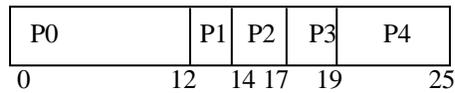

```
| P0          | P1 | P2 | P3 | P4 |
0            12   14  17  19   25
```

**Figure 8: Gantt chart for FCFS**

*b. Shortest Job First*

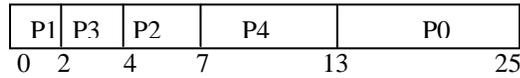

```
| P1 | P3 | P2 |   P4    |    P0    |
0    2    4    7         13         25
```

**Figure 9: Gantt chart for SJF**

*c. Round Robin*
Assign time quantum as 5 ms for each process.

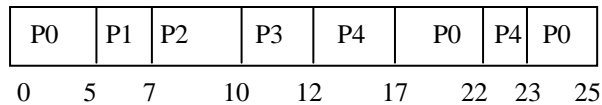

```
| P0 | P1 | P2 | P3 | P4 | P0 | P4 | P0 |
0    5    7    10   12   17   22   23   25
```

**Figure 10: Gantt chart for Round Robin**

*d. Priority Scheduling*
Priority is assigned for each process as follows.

| Process ID | Burst Time(ms) | Priority |
|---|---|---|
| P0 | 12 | 3 |
| P1 | 2 | 1 |
| P2 | 3 | 3 |
| P3 | 2 | 4 |
| P4 | 6 | 2 |

**Table 2: Processes with Its Id, Burst Time and Priority**

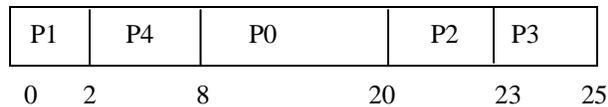

```
| P1 |  P4  |    P0     |   P2   | P3 |
0    2      8            20      23   25
```

**Figure 11 Gantt chart for Priority Scheduling**

For example, turnaround time for the process is calculated as time of submission of a process to the time of completion of the process is obtained through Gantt chart for SJF scheduling. Turnaround time for process P0, P1, P2, P3 & P4 is observed as 25,2,7,4 & 13 respectively and average turnaround time is (25+2+7+4+13)/5=10.2 ms.
The waiting time for the process is calculated as time taken by the process to wait in the ready queue is observed from Gantt chart for SJF scheduling. Waiting time for process P0, P1, P2, P3 & P4 is obtained as 13, 0, 4, 2 & 7 respectively and average waiting time is (13+0+4+2+7)/5=5.2ms. Similarly the turnaround time and waiting time is calculated for all other algorithms and summarized in Table 3 and Table 4.

| Process ID | Turnaround Time (ms) | | | |
|---|---|---|---|---|
| | FCFS | SJF | Round Robin | Priority Scheduling |
| P0 | 12 | 25 | 25 | 20 |
| P1 | 14 | 2 | 7 | 2 |
| P2 | 17 | 7 | 10 | 23 |
| P3 | 19 | 4 | 12 | 25 |
| P4 | 25 | 13 | 23 | 8 |
| Avg. Turnaround Time | 17.4 | 10.2 | 15.4 | 15.6 |

**Table 3: Turnaround Time For Individual Process And Average Turnaround Time For Each Scheduling.**

| Process ID | Waiting Time (ms) | | | |
|---|---|---|---|---|
| | FCFS | SJF | Round Robin | Priority Scheduling |
| P0 | 0 | 13 | 13 | 8 |
| P1 | 12 | 0 | 5 | 0 |
| P2 | 14 | 4 | 7 | 20 |
| P3 | 17 | 2 | 10 | 23 |
| P4 | 19 | 7 | 17 | 2 |
| Avg. Waiting Time | 12.4 | 5.2 | 10.4 | 10.6 |

**Table 4: Waiting Time For Individual Process And Average Waiting Time For Each Scheduling**

From the above discussion it is clear that First Come First Serve (FCFS) & Shortest Job First (SJF) is generally suitable for batch operating systems and Round Robin (RR) & Priority Scheduling (PS) is suitable for time sharing systems. No algorithm is optimum for all type of jobs. Hence it is necessary to develop an algorithm with an optimum criteria and suitable for all scenarios.

## CONCLUSION & FUTURE WORK

The treatment of shortest process in SJF scheduling tends to result in increased waiting time for long processes. And the long process will never get served, though it produces minimum average waiting time and average turnaround time. It is recommended that any kind of simulation for any CPU scheduling algorithm has limited accuracy. The only way to evaluate a scheduling algorithm to code it and has to put it in the operating system, only then a proper working capability of the algorithm can be measured in real time systems.